\documentclass[prl,twocolumn]{revtex4}

\usepackage{amssymb, amsmath, amsfonts, latexsym, verbatim, mathrsfs}

\usepackage{graphicx}

\begin{document}

\title{Comment on ``Is a System's Wave Function in One-to-One Correspondence with Its Elements of Reality?"}
\thanks{This research is supported by the ARO MURI grant W911NF-11-1-0268.}

\author{GianCarlo Ghirardi}
\email{ghirardi@ictp.it}
\affiliation{Department of Physics, University of Trieste, and the Abdus Salam ICTP, Trieste (Italy)}
\author{Raffaele Romano}
\email{rromano@iastate.edu}
\affiliation{Department of Mathematics, Iowa State University, Ames, IA (USA)}

\maketitle


The one-to-one correspondence between quantum state vectors and elements of reality~\cite{colbeck}
follows from the assumptions of validity of
quantum mechanics, $QM$, and a strong request of freedom of choice of measurement settings, $FR$, introduced
in~\cite{colbeck2}. Here we argue that in~\cite{colbeck,colbeck2}
$FR$ is improperly identified with the free will assumption, producing erroneous conclusions
(for a more epistemologically oriented criticism see~\cite{ghirardi}).
In particular, the no-go theorem on $\psi$-epistemic
models presented in~\cite{colbeck} does not have general validity.

We consider two space-like separated observers performing
local measurements on the two parties of an entangled state $\psi$. The
measurement settings are given by vectors $A$ and $B$, the outcomes are denoted
by $X$ and $Y$. Following~\cite{colbeck,colbeck2}, we assume there is additional information on the ontic
state $\lambda$ (the {\it complete specification} of the state of the system, in principle
non completely accessible), obtained through a measurement with setting $C$ and
output $Z$ (we do not exclude the case $Z = \lambda$). We consider all these quantities as random variables.

The $FR$ assumption is the condition that {\it the input $A$ can be chosen to be uncorrelated with all
the space-time random variables whose coordinates lie outside the future light-cone of its coordinates}~\cite{colbeck2},
and the same requirement holds also for $B$ and $C$. This assumption is expressed by the following requests on the
conditional probabilities:
\begin{equation*}
P_{A|BCYZ} = P_A, \quad
P_{B|ACXZ} = P_B, \quad
P_{C|ABXY} = P_C,
\end{equation*}
which are all needed to derive the main results of~\cite{colbeck,colbeck2}.
However, we notice that the free will condition can be consistently expressed in a different form,
by making reference exclusively to the fact that the two observers
can independently choose which observables to measure:
\begin{equation*}
P_{A|B\lambda} = P_A, \quad P_{B|A\lambda} = P_B,
\end{equation*}
where $\lambda$ is the aforementioned ontic state. This condition, denoted by $FR^{\prime}$
in the following, produces the relevant factorization $P_{AB\lambda} = P_A P_B P_{\lambda}$.
Meaningfully, $FR^{\prime}$ is unrelated with the physically
important assumption that the two observers cannot communicate superluminally, denoted as $NS$,
and expressed by $P_{X|AB} = P_{X|A}$ and $P_{Y|AB} = P_{Y|B}$. This is reasonable: one could
have models in which free will and superluminal signalling coexist.
Notice that $FR \Rightarrow NS$, supporting the idea that $FR$ represents more than the free choice assumption.

We observe that in~\cite{colbeck2}
it is pointed out that the information supplementing $\psi$ {\it must be static, that is, its behavior cannot depend on
where or when it is observed}. Otherwise said, the region of events corresponding to the acquisition of this
information can be chosen to be space-like with respect to the events associated to $A$ and $B$, so that $P_{CZ|ABXY} = P_{CZ}$. This statement is presented
as a simple remark in~\cite{colbeck2}; nonetheless,
here we find convenient to consider it as a new assumption, denoted by $ST$. It turns out that
\begin{equation}\label{imply}
    FR^{\prime} \wedge NS \wedge ST \Rightarrow FR.
\end{equation}
In fact, from $ST$ it follows that $P_{ABY|CZ} = P_{ABY}$;
moreover we have
\begin{equation}\label{eq1}
    P_{ABY|CZ} = P_{A|BYCZ} P_{BY|CZ}
                           = P_{A|BYCZ} P_{BY}
\end{equation}
by using again $ST$, but also
\begin{equation}\label{eq2}
    P_{ABY} = P_{AB} P_{Y|AB} =  P_A P_B P_{Y|B}
                           = P_A P_{BY}
\end{equation}
from $NS$ and $FR^{\prime}$. By comparing (\ref{eq1})
and (\ref{eq2}) we find that $P_{A|BYCZ} = P_A$, and a similar argument proves that
$P_{B|AXCZ} = P_B$. Finally, $P_{C|AXBY} = P_C$ is a direct implication of $ST$.

Therefore, violation of $FR$ does not necessarily imply lack of free will
as long as $ST$ or $NS$ are violated. This means that $\psi$-epistemic models fully consistent
with quantum mechanics, with the free will assumption and without superluminal communication are indeed possible,
as long as the supplementary information on the ontic state is not static $P_{C|ABXY} = P_C$.
For instance, they can be easily built by following the lines described in~\cite{lewis}.


\end{document}